\documentclass[10pt,a4paper,aps,preprint]{revtex4-1}
\usepackage[utf8]{inputenc}
\usepackage{color}
\usepackage{float}
\usepackage{amsmath}
\usepackage{amssymb}
\usepackage{graphicx}
\usepackage[unicode=true,pdfusetitle,
 bookmarks=true,bookmarksnumbered=false,bookmarksopen=false,
 breaklinks=false,pdfborder={0 0 1},backref=false,colorlinks=true]
 {hyperref}
\usepackage{breakurl}



\begin{document}

\title{Twisted Supersymmetry in a Deformed Wess-Zumino Model in $\left(2+1\right)$
Dimensions }

\author{C. Palechor}
\email{caanpaip@gmail.com}

\author{A. F. Ferrari}
\email{alysson.ferrari@ufabc.edu.br}

\author{A. G. Quinto}
\email{andres.quinto@ufabc.edu.br}

\affiliation{Universidade Federal do ABC - UFABC, Avenida dos Estados 5001, 0910-580,
Santo André, SP, Brazil.}
\begin{abstract}
Non-anticommutative deformations have been studied in the context
of supersymmetry (SUSY) in three and four space-time dimensions, and
the general picture is that highly nontrivial to deform supersymmetry
in a way that still preserves some of its important properties, both
at the formal algebraic level (e.g., preserving the associativity
of the deformed theory) as well as at the physical level (e.g., maintaining
renormalizability). The Hopf algebra formalism allows the definition
of algebraically consistent deformations of SUSY, but this algebraic
consistency does not guarantee that physical models build upon these
structures will be consistent from the physical point of view. We
will investigate a deformation induced by a Drinfel'd twist of the
${\cal N}=1$ SUSY algebra in three space-time dimensions. The use
of the Hopf algebra formalism allows the construction of deformed
${\cal N}=1$ SUSY algebras that should still preserve a deformed
version of supersymmetry. We will construct the simplest deformed
version of the Wess-Zumino model in this context, but we will show
that despite the consistent algebraic structure, the model in question
is not invariant under SUSY transformation and is not renormalizable.
We will comment on the relation of these results with previous ones
discussed in the literature regarding similar four-dimensional constructions.
\end{abstract}
\maketitle

\section{Introduction}

The non-commutativity of space-time coordinates (henceforth referred
to simply as ``non-commutativity'') was initially proposed as an alternative
to solve the problem of ultraviolet (UV) divergences in quantum electrodynamics\,\cite{HeisenbergtoPeierls(1930)1985}.
The first model that appeared in this context was studied by Snyder
in 1947\,\cite{Snyder1947}, but the idea of non-commutativity was
subdued by the success of renormalization theory to deal with UV divergences.
For some time, the notion of non-commutative manifolds was mostly
developed in the context of mathematics and mathematical physics by
Connes, Woronowics and Drinfel'd\,\cite{Connes2000,Woronowicz1987,Drinfeld},
among others, while some physical applications started to be discussed
in the beginning of this century\,\cite{Jackiw2002}. 

A main motivation for the contemporary interest in the non-commutativity
is linked to the idea that in a quantum theory which incorporates
gravity, the nature of space-time may change in short distances, near
to Planck scales\,\cite{Doplicher1994,Doplicher1995}. In string
theory, the non-commutativity appears in a natural way in the low
energy limit in the presence of a constant background Neveu-Schwarz
two-form $B_{\mu\nu}$\,\cite{Seiberg1999}, leading to an effective
field theory that lives in a space-time where coordinates have nontrivial
commutation relations of the form
\begin{equation}
\left[x^{\mu},x^{\nu}\right]=\left(B^{-1}\right)^{\mu\nu}\thinspace.\label{eq:canonicalNC}
\end{equation}
The simplicity of this particular type of non-commutativity, where
the commutators of space-time coordinates is a constant tensor, allows
the definition of non-commutative versions of known quantum field
models in a way that is very well suited for perturbative calculations.
This approach became known as \emph{canonical non-commutativity},
and it was extensively studied in\,\cite{douglas:2001ba,Szabo:2001kg}.
Supersymmetric models with canonical non-commutativity could be easily
defined since the deformation given in\,\eqref{eq:canonicalNC} does
not interfere with the Grassmanian coordinates of the superspace\,\cite{Terashima:2000xq},
and it was in fact shown that SUSY was an important ingredient to
tame some of the potentially dangerous UV/IR divergences present in
non-commutative models\,\cite{ferrari:2003vs,ferrari:2003ma,Ferrari2004a,Ferrari2004}.
The canonical non-commutativity was even considered in the context
of non-relativistic quantum mechanical models, where several interesting
effects were unveiled\,\cite{gamboa:2000yq,duval:2000xr,chaichian:2000si,nair:2000ii,Ferrari:2010en,Ferrari:2012bv}. 

The canonical non-commutativity, however, involves a preferential
direction in space-time, given by the constant background tensor in\,\eqref{eq:canonicalNC},
inducing an explicit violation of Lorentz invariance. It is an interesting
and nontrivial problem to introduce non-commutativity in space-time
while still preserving Lorentz invariance: one possibility to do this
is by means of the Hopf algebra formalism. In\,\cite{Chaichian2004},
it was shown that although canonical non-commutative theories violate
the Lorentz invariance, they respect another closely related symmetry,
namely the twisted Lorentz symmetry. In the context of Hopf algebras,
the twisted Lorentz symmetry can be understood as a deformation of
the standard Poincaré algebra by means of a Drinfel'd twist\,\cite{Abe1980,Drinfeld,S1995},
such that the Poincaré algebra is not deformed, while the co-algebra
is. The reader can find some studies in the literature regarding Hopf
algebras and deformations associated to Lie algebras in\,\cite{Castro2008,Castro2011,Aschieri2006b},
for example.

Given the interest in the study of theories with non-commutativity
in space-time, it became a natural question to introduce the same
ideas in the superspace, looking for supersymmetric models that live
in deformed superspaces. Besides the already quoted possibility of
the canonical case, in which the supersymmetry structure of the models
is left untouched by the deformation, one may entertain the possibility
of building non-anticomutative (NAC) models, where the algebra of
the Grassmanian coordinates is deformed. General superspace deformations
where considered in\,\cite{Ferrara2000,Klemm2003}, where it was
shown for example that preserving the associativity of the product
of superfields can only be achieved in very specific cases. 

The possibility of NAC deformations was also shown to appear in the
context of superstring theory\,\cite{Seiberg2003}, in which case
the non-anticomutativity of the Grassmanian coordinates is associated
to the presence of a symmetric constant graviphoton field $C^{ab}$,
viz.
\begin{equation}
\left\{ \theta^{a},\theta^{b}\right\} =C^{ab},\thinspace\thinspace\left\{ \bar{\theta}^{\dot{a}},\bar{\theta}^{\dot{b}}\right\} =0\thinspace,\label{eq:conmu}
\end{equation}
where latin indices are indices of two-component spinors. This deformation
is only possible in Euclidean space-times, where $\theta$ and $\bar{\theta}$
are not related by complex conjugation. Besides, the deformation described
by Eq.\,\eqref{eq:conmu} breaks half of the original supersymmetry,
hence this construction became known as $\mathcal{N}=1/2$ SUSY. More
explicitly, one may observe that the only anticommutation relation
between supercharges that are modified by the background tensor $C^{ab}$
is
\begin{equation}
\left\{ \bar{Q}_{a},\bar{Q}_{b}\right\} =-4C^{ab}\sigma_{a\dot{a}}^{\mu}\sigma_{b\dot{b}}^{\nu}\frac{\partial^{2}}{\partial y^{\nu}\partial y^{\mu}}\thinspace.\label{eq:half}
\end{equation}
In practice, the anticommutation relations in Eq.\,\eqref{eq:conmu}
can be introduced in a given supersymmetric action by replacing the
usual product of superfields by the Moyal product
\begin{align}
f\left(\theta\right)\star g\left(\theta\right) & =f\left(\theta\right)\exp\left(-\frac{C^{ab}}{2}\frac{\overleftarrow{\partial}}{\partial\theta^{a}}\frac{\overrightarrow{\partial}}{\partial\theta^{a}}\right)g\left(\theta\right).\label{eq:product}
\end{align}
Using the star product, it is possible to define a deformed Wess-Zumino
(WZ) Lagrangian
\begin{align}
S_{WZ}^{*} & =\int d^{4}\theta\:\Phi\bar{\Phi}+\int d^{2}\theta\,\left(\frac{1}{2}m\Phi\star\Phi+\frac{1}{3}\Phi\star\Phi\star\Phi\right)+\int d^{2}\bar{\theta}\:\left(\frac{1}{2}m\bar{\Phi}\star\bar{\Phi}+\frac{1}{3}\bar{\Phi}\star\bar{\Phi}\star\bar{\Phi}\right)\thinspace.\label{eq:WZ4d}
\end{align}
In this case the supercharge $Q$ generates a symmetry of the theory,
while $\bar{Q}$ does not, showing again that only half of the supersymmetry
is preserved by\,\eqref{eq:conmu}. By expanding the Moyal products
in\,\eqref{eq:WZ4d} in a (finite) power series in $C^{ab}$, and
then integrating in the Grassmann coordinates to obtain the deformed
WZ action in terms of component fields, it can be shown that
\begin{equation}
S_{WZ}^{*}=S_{WZ}-\frac{1}{3}g\thinspace\det C\,\int d^{4}x\thinspace F^{3}\thinspace,\label{eq:WZ4d-2}
\end{equation}
where $F$ is the auxiliary field contained in $\Phi\left(y,\theta\right)$
and $S_{WZ}$ the undeformed WZ action. This result shows that the
NAC in this case amounts to the addition of a single term in the Lagrangian,
proportional to $F^{3}$. 

The study of the quantum properties of the ${\cal N}=1/2$ WZ model
was reported in\,\cite{Grisaru:2003fd,Grisaru:2004qw,Grisaru:2005we},
using the spurion field formalism to include the $F^{3}$ term present
in Eq.\,\eqref{eq:WZ4d-2} within the standard superfield formalism.
Renormalization was shown to be possible but quite nontrivial, since
a finite number of additional counterterms have to be included in
the theory to absorb UV divergences of the quantum effective action. 

\medskip{}

In $\left(2+1\right)$ dimensions, the structure of supersymmetric
models is simpler in a sense, since the Grassmann coordinates $\theta$
are real (due to the Lorentz group being related to $SL(2,\mathbb{R})$
instead of $SL(2,\mathbb{C})$ as in $\left(3+1\right)$ dimensions),
and the notion of chirality is absent. This simplicity, however, seems
to leave less room to define deformations of supersymmetry that still
preserve at least some of the interesting properties of supersymmetric
models. Some options were first discussed in\,\cite{Ferrari2006},
where the strategy was to start with ${\cal N}=2$ SUSY, which was
brought down to ${\cal N}=1$ by the deformation. The problem of deformation
of ${\cal N}=2$ three-dimensional SUSY have been revisited recently\,\cite{Faizal:2011en,Faizal:2012zi,Gama:2014hwa},
mostly motivated by the search of deformed variants of ABJM theories\,\cite{Aharony:2008ug}.

In this paper, we want to investigate one alternative way to introduce
NAC in three-dimensional SUSY that has not yet been developed, which
is the use of twisted symmetries. In the case of four-dimensional
SUSY, there was some extensive work investigating several possible
deformations\,\cite{Ihl:2005zd,Irisawa:2006xx,kobayashi:2004ep}.
However, it was already pointed out that algebraic consistency is
not enough to ensure the construction of physically meaningful models.
In\,\cite{Dimitrijevic:2007cu,Dimitrijevic:2009mt,Dimitrijevic:2010yv,Dimitrijevic:2011zg},
two different twist deformations of four-dimensional SUSY were throughly
examined, and it was found that one may write actions in terms of
superfields that, despite being formally covariant, are actually not
invariant under SUSY transformations. Even if this problem can be
circumvented in order do define a deformed WZ model at the classical
level, investigation of the quantum corrections showed that it turn
out to be non-renormalizable. The need to preserve the notion of chirality
(by means of the introduction of non-linear projection operators)
was at the core of the problems reported in\,\,\cite{Dimitrijevic:2007cu,Dimitrijevic:2009mt,Dimitrijevic:2010yv,Dimitrijevic:2011zg},
so one might wonder whether the situation in three-dimensional case,
where there is no chirality, could be better.

The formalism of twist deformations would be particularly interesting
in three dimensions because it could potentially implement NAC in
an ${\cal N}=1$ supersymmetric model, since on the Hopf algebraic
description, one may deform only the co-algebra and not the algebra
itself. An undeformed SUSY algebra would mean, in principle, a NAC
model invariant under the same SUSY transformations as the undeformed
one, but with a deformed Leibnitz rule when the supercharges act on
the product of superfields. However, we will show that the same problems
found in four-dimensional models appear here: the component version
of the most natural definition of a deformed WZ model fails to be
supersymmetric invariant, and when quantum corrections are calculated
using the superfield formalism, the model turns out to be non-renormalizable.
The end result is that SUSY in $\left(2+1\right)$ dimensions, despite
being somehow simpler in structure, seems to impose stringent restrictions
on the possible NAC deformations one can consistently define, even
when the algebraic machinery of Hopf algebras is used. \medskip{}

This paper is organized as follows. In section\,\ref{HopfAlgebra},
we give a brief review of SUSY in the language of Hopf algebras (for
a more detailed review, see the discussion of the four-dimensional
case in\,\cite{kobayashi:2004ep,Dimitrijevic:2007cu}, for example).
The deformation of the supersymmetry algebra using a twist element
is introduced in section\,\ref{TwistDeformation}. In section\,\ref{TwistDeformation}
we define a covariant field theory under the deformed algebra, which
would be the deformed version of the WZ model. We discuss the basic
properties of this model, showing how despite being written in terms
of covariant superfields and operators, the action fails to be invariant
under component supersymmetry transformations. Regardless this issue,
the model can be quantized and its renormalization studied, by using
the spurion technique, and in section\,\ref{QuantumProperties} we
show that the models turns out to be non-renormalizable. Section\,\ref{sec:Concluding-Remarks}
contains our conclusions. The notations and conventions of\,\cite{Gates2001}
are used throughout the text.

\section{\label{HopfAlgebra}Supersymmetry in the Hopf Algebra Formalism}

The supersymmetric Poincaré superalgebra has an universal enveloping
algebra which has the natural structure of a Hopf superalgebra. This
construction preserves the main properties of the usual SUSY algebra,
such as the anticommutators and the Jacobi identities\,\cite{Zhang1999}.
Indeed, the SUSY algebra in $\left(2+1\right)$ dimensions, which
will be denoted by $\mathcal{SP}$ is given by,\begin{subequations}\label{eq:The-supersymmetry-algebra}
\begin{align}
\left[P_{ab},P_{cd}\right] & =0\,,\\
\left\{ Q_{a},Q_{b}\right\}  & =2P_{ab}\,,\\
\left[Q_{a},P_{cd}\right] & =0\,.
\end{align}
\end{subequations}Here, $P_{ab}$ is the generator of space-time
translations, represented as a bispinor. The universal enveloping
algebra $\mathcal{U}(\mathcal{SP})$ is defined as the quotient of
the sum of tensor products of the original algebra $\mathcal{SP}$
by the ideal generated by the (anti)commutations relations\,\eqref{eq:The-supersymmetry-algebra}.
The Hopf algebra structure of $\mathcal{U}(\mathcal{SP})$ is encoded
by the coproduct, product and antipode given by
\begin{align}
\Delta(\zeta) & \equiv\sum_{i}(\zeta_{1})_{i}\otimes(\zeta_{2})_{i}=\zeta\otimes1+1\otimes\zeta,\label{usualcoproduct}\\
\mu(\zeta\otimes\eta) & =\zeta\cdot\eta,\\
S(\zeta) & =-\zeta,\,S(1)=1,\eta
\end{align}
where in Eq.\,\eqref{usualcoproduct} we have used the Sweedler notation\,\cite{Sweedler1969},
and $\zeta,\eta\in\mathcal{U}(\mathcal{SP})$. The algebraic structure\,\eqref{eq:The-supersymmetry-algebra}
is encoded in the adjoint action of one operator $\zeta$ into another
$\eta$, i.e., 
\begin{align}
ad_{\zeta} & =(-1)^{\kappa(\eta)\,\kappa(\zeta_{{\scriptscriptstyle 2}})}\zeta_{1}\cdot\eta\cdot S\left(\zeta_{2}\right)\nonumber \\
 & =\zeta\cdot\eta-(-1)^{\kappa(\eta)\,\kappa(\zeta)}\,\eta\cdot\zeta\thinspace,\label{eq:adjoint}
\end{align}
where $\kappa$ is the usual parity function.

The superalgebra $\mathcal{U}(\mathcal{SP})$ act on superfields,
which are functions of the superspace with coordinates $z=\left(x^{ab},\theta^{a}\right)$
satisfying
\begin{align}
\left[x^{mn},x^{rs}\right] & =\left[x^{mn},\theta^{a}\right]=0\thinspace,\\
\left\{ \theta^{a},\theta^{b}\right\}  & =0\thinspace.
\end{align}
This action can be represented by first order differential operators
as follows,\begin{subequations}\label{eq:generators}
\begin{align}
Q_{a} & =i\left(\partial_{a}-i\,\theta^{b}\,\partial_{ba}\right)\thinspace,\\
P_{ab} & =i\,\partial_{ab}\thinspace,\\
D_{a} & =\partial_{a}+i\,\theta^{b}\partial_{ba}\,,
\end{align}
\end{subequations} where $D_{a}$ are the supercovariant derivatives,
which are essential in the definition of covariant supersymmetric
actions.

The superfield themselves encompass an algebra with product $m$,
which in the standard (undeformed) case is given by the pointwise
product
\begin{equation}
m\left(\Phi\otimes\Psi\right)=\Phi\left(z\right)\cdot\Psi\left(z\right)\thinspace.
\end{equation}
In the Hopf algebra formalism, the Leibniz rule is represented by
the covariant action of the Hopf algebra on the algebra of the superfields,
i.e.,
\begin{equation}
\zeta\vartriangleright(m\left(\Phi\otimes\Psi\right))=m\left(\Delta(\zeta)\vartriangleright(\Phi\otimes\Psi)\right),
\end{equation}
which, for the undeformed coproduct\,\eqref{usualcoproduct}, reduces
the usual Leibnitz rule,
\begin{align}
\zeta(\Phi\cdot\Psi) & =\zeta(\Phi)\cdot\Psi+(-1)^{\kappa(\zeta)\kappa(\Psi)}\Phi\cdot\zeta(\Phi)\,.\label{leibniz}
\end{align}

Superfields can be decomposed in terms of component fields. For the
simplest case of a scalar superfield $\Phi\left(x,\theta\right)$,
we have
\begin{equation}
\Phi\left(x,\theta\right)=A\left(x\right)+\theta^{a}\psi_{a}\left(x\right)-\theta^{2}F\left(x\right)\,,\label{eq:Phicomponents}
\end{equation}
where $A$ and $\psi$ are scalar and spinorial fields, and $F$ is
an auxiliary field. SUSY transformation are generated by the supercharges
$Q$,
\begin{align}
\delta_{\xi}\Phi\left(x,\theta\right) & \equiv i\,\xi^{a}\,Q_{a}\Phi\left(x,\theta\right),
\end{align}
which in terms of component fields amounts to,
\begin{align}
\delta A\left(x\right) & =-\xi^{a}\psi_{a}\left(x\right)\,,\\
\delta\psi_{a}\left(x\right) & =-\xi^{b}\left(\epsilon_{ab}\,F\left(x\right)+i\,\partial_{ab}\,A\left(x\right)\right)\,,\\
\delta F\left(x\right) & =-i\,\xi^{a}\,\partial_{a}^{\,\,b}\,\psi_{b}\left(x\right)\,.
\end{align}
In the undeformed case, $Q$ is represented by a first order differential
operator, which satisfies the standard Leibnitz rule when applied
on product of superfields,
\begin{equation}
\delta_{\xi}\left(\Phi\cdot\Psi\right)=\delta_{\xi}\left(\Phi\right)\cdot\Psi+\Phi\cdot\delta_{\xi}\left(\Psi\right)\thinspace,
\end{equation}
which, in the Hopf algebra formalism, is encoded by the standard coproduct\,\eqref{usualcoproduct}.

\section{\label{TwistDeformation}Twist deformation of the SUSY algebra}

In the Hopf algebra formalism, the deformation can be introduced as
a Drinfel'd twist\,\cite{Drinfeld,majidbook:1998rq,Aschieri2006}.
One starts by choosing a twist element, which we postulate is given
by\begin{subequations}\label{eq:TwistElement}
\begin{align}
\mathcal{F} & =f^{a}\otimes f_{a}=\exp\left[\frac{1}{2}\,C^{ab}\:\partial_{a}\otimes\partial_{b}\right],\\
\mathcal{F}^{-1} & =\overline{f}^{a}\otimes\bar{f}_{a}=\exp\left[-\frac{1}{2}\,C^{ab}\:\partial_{a}\otimes\partial_{b}\right],
\end{align}
\end{subequations}where $C^{ab}$ is a symmetric matrix\footnote{One clarification is in order here: to use this particular twist,
one has to enlarge the superalgebra $\mathcal{U}(\mathcal{SP})$ by
including the Grassmanian derivatives $\partial_{a}$ as new generators,
in equal footing to $Q_{a}$, $D_{a}$ and $P_{ab}$. This can be
done without spoiling the algebraic consistency of the construction.}. The twist element\,\eqref{eq:TwistElement} can be shown to satisfy
the 2-cocycle condition
\begin{equation}
\mathcal{F}\left(\Delta\otimes id\right)\mathcal{F}=\mathcal{F}\left(id\otimes\Delta\right)\mathcal{F}\thinspace,
\end{equation}
which guarantees associativity of the construction. This is similar
to the twist considered in\,\cite{Dimitrijevic:2007cu,Dimitrijevic:2011zg}. 

Since the Grassmanian derivatives are nilpotent and anticommutative,
when expanded in powers of $C$ we find $\mathcal{F}$ to be finite,
\begin{equation}
\mathcal{F}=1\otimes1+\frac{1}{2}\:C^{ab}\,\partial_{a}\otimes\partial_{b}-\frac{1}{8}\,C^{ab}C^{mn}\:\partial_{a}\,\partial_{m}\otimes\partial_{b}\,\partial_{n},
\end{equation}
such as in the ${\cal N}=1/2$ SUSY\,\cite{Seiberg2003}, and in
the twisted supersymmetry model studied in\,\cite{Dimitrijevic:2007cu,Dimitrijevic:2011zg},
both in four space-time dimensions, but differently from the three-dimensional
deformation considered in\,\cite{Ferrari2006}, in which the expansion
of the Moyal product has infinite terms. It is also interesting to
stress that, differently from what happens in four dimensions, a similar
twist involving the supercovariant derivative $D_{\alpha}$ instead
of $Q_{\alpha}$ would not be finite, since the $D$'s do not anticommute
among themselves, so we do not consider a \char`\"{}D-deformation\char`\"{}
as in\,\cite{Dimitrijevic:2009mt,Dimitrijevic:2010yv}, which would
be much more complicated in our case. 

The deformation is implemented on the algebra of superfields by means
of the deformed star product given by
\begin{align}
\Phi\left(z\right)\star\Psi\left(z\right) & =m^{\mathcal{F}}(\Phi\otimes\Psi)=m(\mathcal{F}^{-1}\vartriangleright\left(\Phi\otimes\Psi\right))\nonumber \\
 & =(-1)^{\kappa(\Phi)\,\kappa(\bar{f}_{a})}\:\left(\bar{f}^{a}\triangleright\Phi\right)\cdot\left(\bar{f}_{a}\vartriangleright\Psi\right)\nonumber \\
 & =\Phi\cdot\Psi-\frac{1}{2}(-1)^{\kappa(\Phi)}\:C^{ab}\,\partial_{a}\Phi\cdot\partial_{b}\Psi-\frac{1}{8}\,C^{ab}C^{mn}\:\partial_{a}\,\partial_{m}\Phi\cdot\partial_{b}\,\partial_{n}\Psi\,.\label{Star product}
\end{align}
The start product introduce NAC in the superspace, since
\begin{equation}
\left[x^{mn}\:\overset{\star}{,}\:x^{rs}\right]=\left[x^{mn}\:\overset{\star}{,}\;\theta^{a}\right]=0\,,\thinspace\left\{ \theta^{a}\:\overset{\star}{,}\:\theta^{b}\right\} =C^{ab}\,,\label{Non commutativity}
\end{equation}
where these $\star$ (anti)commutators are defined by replacing the
usual product of functions by the star product. 

In the context of ${\cal N}=1/2$ SUSY, the deformation of the Poincaré
superalgebra is obtained by formally calculating the anticommutators
of the generators\,\eqref{eq:generators}, taking into account Eq.\,\eqref{Non commutativity}.
This would lead to 
\begin{align}
\left\{ Q_{a},Q_{b}\right\}  & _{\star}=2P_{ab}-C^{mn}P_{ma}P_{nb}\,,\label{Q commutator}\\
\left\{ D_{a},D_{b}\right\}  & _{\star}=2P_{ab}+C^{mn}P_{ma}P_{nb}\,,\\
\left\{ Q_{a},D_{b}\right\}  & _{\star}=-i\,C^{mn}P_{ma}P_{nb}\,,
\end{align}
where the breaking of SUSY becomes manifest. However, in the Hopf
algebra formalism, this is actually not correct since the star product
is, at this point, defined only for superfields, and not for operators.
To properly define the star (anti)commutators, we extend the discussion
presented in\,\cite{castro:2008su} for the case of graded superalgebras.
We define the star product between two elements of the superalgebra
$\mathcal{U}(\mathcal{SP})$ as
\begin{align}
\zeta\,\underline{\star}\,\eta & =\sum_{a}(-1)^{\kappa(\bar{f}_{a})\,\kappa(u)}\:\bar{f}^{a}\left(\zeta\right)\cdot\bar{f}_{a}\left(\eta\right),
\end{align}
where $\bar{f}^{a}\left(\zeta\right)\equiv ad_{\bar{f}^{a}}(\zeta)$,
the adjoint action defined in\,\eqref{eq:adjoint}. With this definition,
one can calculate for example
\begin{align}
\left\{ Q_{a}\:\overset{\underline{\star}}{,}\:Q_{b}\right\}  & =Q_{a}\,\underline{\star}\,Q_{b}+Q_{b}\,\underline{\star}\,Q_{a}\nonumber \\
 & =2P_{ab}-C^{mn}P_{ma}P_{nb}\,,
\end{align}
reproducing the result in Eq.\,\eqref{Q commutator}. The same can
be done for the other generators.

In the context of twisted deformations, for any of the operators $\eta$
that generate $\mathcal{U}(\mathcal{SP})$, one may associate a new
(deformed) generator $\widetilde{\eta}$, 
\begin{align}
\widetilde{\eta} & =\sum_{a}(-1)^{\kappa(\bar{f}_{a})\,\kappa(\eta)}\:\bar{f}^{a}\cdot\eta\cdot S\left(\bar{f}_{a}\right),
\end{align}
also belonging to $\mathcal{U}(\mathcal{SP})$, in such a way that
$\widetilde{\eta}$ satisfies the original, undeformed algebra\,\cite{castro:2008su}.
This is a clear advantage of this formalism. For example, considering
the supersymmetric generator $Q_{a}$, we have 
\begin{align}
\widetilde{Q}_{a}= & Q_{a}+\frac{i}{2}\:C^{lm}\:\partial_{m}P_{al}\,,
\end{align}
and one can verify that
\begin{align}
\left\{ \widetilde{Q}_{a}\:\overset{\underline{\star}}{,}\:\widetilde{Q}_{b}\right\}  & =2P_{ab}\,.
\end{align}
Clearly, $\widetilde{Q}_{a}$ is not linear in the generators of the
algebra, so indeed it fits only within the Hopf algebraic machinery,
and not the usual Lie algebra formalism. If we consider $\widetilde{Q}_{a}$
as the generator of supersymmetry transformations, we can say we constructed
a deformed NAC superspace, while still preserving supersymmetry.

It is interesting to point out that the possibility of defining nonlinear
generators which satisfy the undeformed algebra was already briefly
pointed out in different contexts\,\cite{Seiberg2003,Ferrari2006}.
In\,\cite{Ferrari2006}, for example, it was considered an ${\cal N}=2$
three-dimensional superspace with real Grassmanian coordinates $\theta_{a}^{1,2}$,
and the deformed algebra
\begin{equation}
\left\{ \theta_{a}^{1},\theta_{b}^{1}\right\} =0,\thinspace\left\{ \theta_{a}^{2},\theta_{b}^{2}\right\} =\Sigma_{ab}\thinspace,
\end{equation}
where $\Sigma_{ab}$ is the deformation parameter. In this formalism,
the SUSY transformation generated by $Q_{a}^{1}$ is preserved, while
the one generated by $Q_{a}^{2}$ is lost, as can be seen by inspecting
their anticommutation relations,
\begin{equation}
\left\{ Q_{a}^{1},Q_{b}^{1}\right\} =2P_{ab},\quad\left\{ Q_{a}^{1},Q_{b}^{2}\right\} =0,\quad\left\{ Q_{a}^{2},Q_{b}^{2}\right\} =2P_{ab}+\Sigma^{cd}P_{ac}P_{bd}\thinspace.
\end{equation}
One may however define the nonlinear generators 
\begin{equation}
\tilde{Q}_{a}^{2}=Q_{a}^{2}+\frac{i}{2}\,\Sigma^{bc}\,\partial_{b}^{2}\,P_{ca}\thinspace,\label{eq:oldtildeQ}
\end{equation}
which satisfy the usual (undeformed) supersymmetry algebra,
\begin{equation}
\left\{ \tilde{Q}_{a}^{2},\tilde{Q}_{b}^{2}\right\} =2P_{ab}\thinspace,
\end{equation}
but this possibility was not fully developed in\,\cite{Ferrari2006}
insomuch as those new generators were represented by nonlinear operators.
We see that the deformed generators given in Eq\,\eqref{eq:oldtildeQ}
fit naturally within the Hopf algebra formalism, realizing an instance
of twisted supersymmetry. 

The coproduct for the deformed generators is defined by
\begin{align}
\Delta_{\star}(\widetilde{\eta}) & =\mathcal{F}\left(\widetilde{\eta}\otimes1+1\otimes\widetilde{\eta}\right)\mathcal{F}^{-1\text{ }},\label{Coproduct}
\end{align}
which is compatible with the star product in Eq.\,\eqref{Star product},
meaning that
\[
\widetilde{\eta}\vartriangleright(m^{\mathcal{F}}\left(\Phi\otimes\Psi\right))=m^{\mathcal{F}}\left(\Delta_{\star}(\widetilde{\eta})\vartriangleright(\Phi\otimes\Psi)\right)\,.
\]

The action of the deformed generators on a single superfield can be,
in a slight abuse of notation, defined as
\begin{equation}
\widetilde{\eta}\vartriangleright\Phi=\eta\vartriangleright\Phi\,,\label{eq:deformedaction}
\end{equation}
meaning the action of the operator $\widetilde{\eta}$ mimics that
of the undeformed generator $\eta$ when acting on a single superfield\,\cite{Aschieri2006,Dimitrijevic:2007cu}.
The action of $\widetilde{\eta}$ on a product of superfields is deformed
according to the coproduct\,\eqref{Coproduct}. For the supersymmetry
generators, this means that the SUSY transformation of a single superfield
is undeformed, 
\begin{equation}
\delta_{\xi}^{\star}\Phi\left(x,\theta\right)\equiv i\xi^{a}\widetilde{Q}_{a}\vartriangleright\Phi\left(x,\theta\right)=i\xi^{a}Q_{a}\Phi\left(x,\theta\right)\,.\label{eq:defsusytrans}
\end{equation}
However, due to the deformed coproduct given in Eq.\,\eqref{Coproduct},
the SUSY transformation of a star product of superfields is modified
according to 
\begin{equation}
\delta_{\xi}^{\star}(\Phi\star\text{\ensuremath{\Psi}})=\left(\delta_{\xi}^{\star}\Phi\right)\star\Psi+\Phi\star\left(\delta_{\xi}^{\star}\Psi\right)+\frac{i}{2}C^{mn}\xi^{a}\,\Bigl(\partial_{m}\Phi\star\partial_{na}\Psi-\left(-1\right)^{\kappa(\Phi)}\partial_{ma}\Phi\star\partial_{n}\Psi\Bigr)\thinspace.\label{deformed_transformation}
\end{equation}
In essence, in this formalism the effect of the deformation is to
modify the Leibnitz rule according to which supercharges act on products
of superfields, while the algebra of supercharges itself is not modified.
This formalism should allows us to define actions involving superfields
that are, in principle, covariant under SUSY transformations. As we
will show in the next section, however, this does not guarantee the
actual SUSY invariance of the model when projected to the physical
(component) fields.

\section{\label{Deformed-Wess-Zumino-action}Deformed Wess-Zumino action}

We consider the deformed WZ action in $\left(2+1\right)$ dimensions,
\begin{align}
\mathcal{S}^{\star}= & -\frac{1}{4}\,\int d^{5}z\,\left(\widetilde{D}^{b}\triangleright\Phi\right)\star\left(\widetilde{D}_{b}\triangleright\Phi\right)\nonumber \\
 & +\frac{1}{2}\int d^{5}z\,m\Phi\star\Phi+\frac{\lambda}{6}\int d^{5}z\,\Phi\star\Phi\star\Phi\,,\label{classical_action}
\end{align}
where $\int d^{5}z\equiv\int\,d^{3}x\,d^{2}\theta\thinspace.$ This
is the standard WZ action, with the usual products replaced by the
star products defined in Eq.\,\eqref{Star product}. Using integration
by parts, one may show from Eq.\,\eqref{Star product} that
\begin{equation}
\int d^{5}z\thinspace{\cal H}\star{\cal G}=\int d^{5}z\thinspace{\cal H}{\cal G}\thinspace,\label{eq:starquadratic}
\end{equation}
where ${\cal H},\thinspace{\cal G}$ are arbitrary superfields. From
this property, together with Eq.\,\eqref{eq:deformedaction}, we
can show that the quadratic term in Eq.\,\eqref{classical_action}
remains undeformed, 
\begin{equation}
\mathcal{S}_{kin}^{\star}=\frac{1}{2}\int d^{5}z\thinspace\left[\Phi D^{2}\Phi+m\Phi^{2}\right]\thinspace.
\end{equation}

As for the cubic interaction terms they reduce to the usual WZ interactions
together with an additional term, 
\begin{equation}
\mathcal{S}_{I}^{\star}=\frac{\lambda}{6}\int d^{5}z\,\Phi\star\Phi\star\Phi=\frac{\lambda}{6}\int d^{5}z\,\Phi\Phi\Phi+\frac{\lambda}{48}\int d^{5}z\,C^{lm}C^{nk}\,\partial^{2}\Phi\partial_{l}\partial_{n}\Phi\partial_{k}\partial_{m}\Phi\thinspace,
\end{equation}
or, in terms of components fields,
\begin{align}
\mathcal{S}_{I}^{\star} & =\int d^{3}x\,\mathcal{L}_{I}+\frac{\lambda}{24}\int d^{3}x\,C^{2}\,F^{3}\thinspace,\label{eq:F3term}
\end{align}
where
\begin{equation}
\det C=C^{2}=\frac{1}{2}\,C^{ml}C^{nk}\,\epsilon_{ln}\epsilon_{mk}\thinspace,\label{eq:detC}
\end{equation}
and $\mathcal{L}_{I}$ is the usual WZ interaction Lagrangian. The
situation here is similar to the case of ${\cal N}=1/2$ SUSY\,\cite{Seiberg2003},
where also the final effect of the deformation in the WZ model is
the addition of a single interaction term involving the auxiliary
field $F$ in the action. 

Despite our construction being formally covariant, to ensure the physical
consistency we have to project the superfield action in term of the
(physical) component fields, and verify explicitly the SUSY invariance.
We start with the quadratic terms in Eq.\,\eqref{classical_action},
and using Eq.\,\eqref{deformed_transformation}, we can write
\begin{align}
\delta_{\xi}^{\star}(\widetilde{D}^{b}\triangleright\Phi\star\widetilde{D}_{b}\triangleright\Phi) & =\delta_{\xi}^{\star}\left(\widetilde{D}^{b}\triangleright\Phi\right)\star\widetilde{D}_{b}\triangleright\Phi+\widetilde{D}^{b}\triangleright\Phi\star\delta_{\xi}^{\star}\left(\widetilde{D}_{b}\triangleright\Phi\right)+\nonumber \\
 & +\frac{i}{2}C^{mn}\xi^{a}\,\Bigl(\partial_{m}\left(\widetilde{D}^{b}\triangleright\Phi\right)\star\partial_{na}\left(\widetilde{D}_{b}\triangleright\Phi\right)+\partial_{ma}\left(\widetilde{D}^{b}\triangleright\Phi\right)\star\partial_{n}\left(\widetilde{D}_{b}\triangleright\Phi\right)\Bigr)\thinspace.
\end{align}
Because this expression is integrated we can use the property\,\eqref{eq:starquadratic},
together with Eq.\,\eqref{eq:deformedaction}, to obtain
\begin{align}
\delta_{\xi}^{\star}(\widetilde{D}^{b}\triangleright\Phi\star\widetilde{D}_{b}\triangleright\Phi) & =\delta_{\xi}^{\star}\left(D^{b}\Phi\right)\cdot D_{b}\Phi+D^{b}\Phi\cdot\delta_{\xi}^{\star}\left(D_{b}\Phi\right)+\nonumber \\
 & +\frac{i}{2}C^{mn}\xi^{a}\,\Bigl(\partial_{m}D^{b}\Phi\cdot\partial_{na}D_{b}\Phi+\partial_{ma}D^{b}\Phi\cdot\partial_{n}D_{b}\Phi\Bigr)\thinspace.
\end{align}
It is easy to verify that the two terms in the second line of the
previous equation cancel among each other, therefore, 
\begin{align}
\delta_{\xi}^{\star}(\widetilde{D}^{b}\triangleright\Phi\star\widetilde{D}_{b}\triangleright\Phi) & =\delta_{\xi}^{\star}\left(D^{b}\Phi\right)\cdot D_{b}\Phi+D^{b}\Phi\cdot\delta_{\xi}^{\star}\left(D_{b}\Phi\right)\nonumber \\
 & =2\delta_{\xi}\left(D^{b}\Phi\right)\cdot D_{b}\Phi\thinspace,
\end{align}
where Eq.\,\eqref{eq:defsusytrans} was used in the last line. This
final expression is clearly SUSY invariant. The same procedure can
be used to argue for the SUSY invariance of the remaining quadratic
term $\Phi\star\Phi$. 

However, upon explicit calculation, the additional interaction term
in Eq,\,\eqref{eq:F3term} is not invariant under the deformed SUSY
transformation. To verify that, we remember that the integration over
Grassmanian coordinates amounts to projecting the last component,
proportional to $\theta^{2}$, of the integrand. So, SUSY invariance
of Eq.\,\eqref{eq:F3term} means that the $\theta^{2}$ component
of $\delta_{\xi}^{\star}\biggl(\Phi\star\Phi\star\Phi\biggr)$ should
be at the most a surface term. The action of $\delta_{\xi}^{\star}$
is distributed among the factors in the star product by the deformed
Leibnitz rule, Eq.\,\eqref{deformed_transformation}. Using also
Eq.\,\eqref{eq:starquadratic}, we have{\small{}
\begin{align}
\delta_{\xi}^{\star}\biggl(\Phi\star\Phi\star\Phi\biggr) & =\left(\delta_{\xi}^{\star}\Phi\right)\star\Phi\star\Phi+\Phi\star\left(\delta_{\xi}^{\star}\Phi\right)\star\Phi+\Phi\star\Phi\star\left(\delta_{\xi}^{\star}\Phi\right)\nonumber \\
 & +\frac{i}{2}C^{mn}\xi^{a}\,\Phi\star\Bigl(\partial_{m}\Phi\star\partial_{na}\Phi-\partial_{ma}\Phi\star\partial_{n}\Phi\Bigr)\nonumber \\
 & +\frac{i}{2}C^{mn}\xi^{a}\,\Bigl(\partial_{m}\Phi\star\partial_{na}\left(\Phi\star\Phi\right)-\partial_{ma}\Phi\star\partial_{n}\left(\Phi\star\Phi\right)\Bigr)\thinspace.\label{eq:55}
\end{align}
}Also because of\,\eqref{eq:starquadratic}, all the terms in the
first line of this last equation are equal. For the second and third
lines, explicit expansion of the star products yields{\small{}
\begin{align}
\frac{i}{2}C^{mn}\xi^{a}\,\Phi\star & \Bigl(\partial_{m}\Phi\star\partial_{na}\Phi-\partial_{ma}\Phi\star\partial_{n}\Phi\Bigr)=\nonumber \\
 & =\frac{i}{2}C^{mn}\xi^{a}\:C^{pq}\left(\Phi\cdot\partial_{p}\partial_{m}\Phi\cdot\partial_{q}\partial_{na}\Phi\right)\thinspace,
\end{align}
}while the third line vanishes. Therefore
\begin{align}
\delta\biggl(\Phi\star\Phi\star\Phi\biggr) & \biggl|_{\theta^{2}}=3\left(\delta_{\xi}^{\star}\Phi\right)\star\Phi\star\Phi\biggl|_{\theta^{2}}+\nonumber \\
 & +\frac{i}{2}C^{mn}\xi^{a}\:C^{pq}\left(\Phi\cdot\partial_{p}\partial_{m}\Phi\cdot\partial_{q}\partial_{na}\Phi\right)\biggl|_{\theta^{2}}\thinspace.
\end{align}
Finally, using Eqs.\,\eqref{Star product}, \eqref{eq:defsusytrans}
and \eqref{eq:Phicomponents}, projecting out only the terms proportional
to $\theta^{2}$, after some algebraic manipulations, we arrive at
\begin{align}
\delta\biggl(\Phi\star\Phi\star\Phi\biggr) & \biggl|_{\theta^{2}}=\left[-\frac{3}{4}i\,C^{2}\,\xi^{a}\partial_{a}^{\,\,b}\psi_{b}\cdot F^{2}\right]+\nonumber \\
 & +\frac{i}{4}C^{mn}C^{pq}\:\epsilon_{pm}\xi^{a}\left[2\:\partial_{na}\psi_{q}\cdot F^{2}-\psi_{q}\partial_{na}F^{2}\right]
\end{align}
which is not a surface term. For this reason the model is not invariant
under deformed SUSY.

This lack of SUSY invariance is a surprise in this formalism since,
differently from the ${\cal N}=1/2$ case in $\left(3+1\right)$ dimensions\,\cite{Seiberg2003},
or also the three-dimensional deformations studied in\,\cite{Ferrari2006,Gama:2014hwa},
in the twist deformation we are considering the deformation is introduced
in a way that the covariance of superfields and the algebra of supercharges
is not deformed. However, this lack of SUSY invariance in the component
formulation, despite the formally covariant construction was already
pointed out in $\left(3+1\right)$ dimensions in\,\cite{Dimitrijevic:2007cu,Dimitrijevic2009,Dimitrijevic2011,Dimitrijevic:2010yv},
where the twist formalism was also used. There, the lack of invariance
was attributed to the need of introducing non-local projection operators
to maintain the notion of chirality. Indeed, in the four-dimensional
case, one has to apply (anti)chiral projectors,
\begin{equation}
P_{1}=\frac{1}{16}\frac{D^{2}\bar{D}^{2}}{\square}\thinspace,\thinspace P_{2}=\frac{1}{16}\frac{\bar{D}^{2}D^{2}}{\square}\thinspace,
\end{equation}
to star products of (anti)chirals superfields to maintain (anti)chirality.
However, this introduces an ambiguity in the definition of the deformed
trilinear interaction term, since for example both $P_{2}\left(\Phi\star\Phi\star\Phi\right)$
and $P_{2}\left(\Phi\star P_{2}\left(\Phi\star\Phi\right)\right)$
are acceptable, however upon explicit calculation, it was shown that
the first expression is not SUSY invariant, while the second is. This
is rather surprising since both are formally covariant. 

We find that, even in the absence of non-local projection operators,
the formal covariance of the superfield action is not enough to guarantee
the SUSY invariance of the model. This seems to be a rather important
shortcoming of the Hopf algebra formalism in the definition of consistent
deformed superfield theories.

\section{\label{QuantumProperties}Quantum Properties and renormalization}

\begin{figure}[h]
\begin{centering}
\includegraphics[scale=0.5]{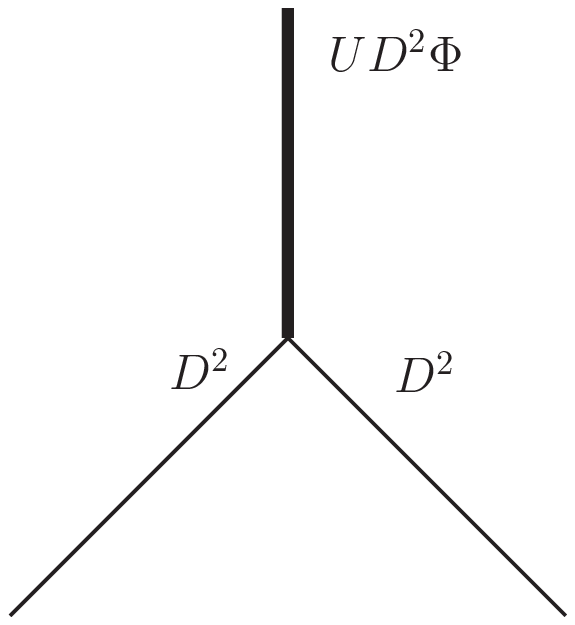} \hspace{2cm}\includegraphics[scale=0.5]{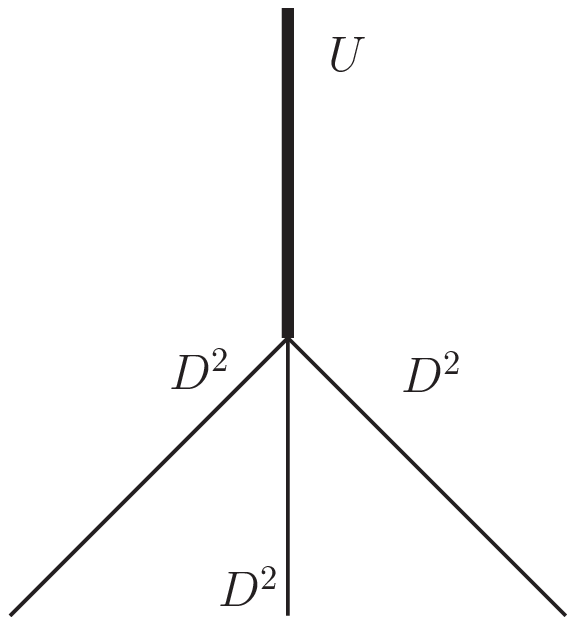}
\par\end{centering}
\caption{\label{fig:New-vertices}New vertices arising from the $U$-term.
Thin lines represent the quantum field, which appears only in internal
lines of the diagrams.}
\end{figure}

Despite the problem with SUSY invariance unveiled in the previous
section, one may still wonder about the quantum properties of the
model defined by Eq.\,\eqref{classical_action}. Indeed, this model
might still define a consistent, yet non-supersymmetric, quantum field
theory. Functional quantization of Eq.\,\eqref{classical_action}
is possible by means of the methods similar to the ones used in the
context of ${\cal N}=1/2$ models in\,\cite{Grisaru:2003fd,Grisaru:2004qw}:
the NAC deformation amounts to the addition of the single (non-invariant)
term in the action, and this could be incorporated in the superfield
formalism by means of a spurion field given by 
\begin{equation}
U\left(z\right)=-\det C\thinspace\theta^{2}\thinspace,
\end{equation}
where we used the definition\,\eqref{eq:detC}. In this way, we may
rewrite Eq.\,\eqref{eq:F3term} as follows,
\begin{align}
\mathcal{S}^{\star} & =\int d^{3}x\,d^{2}\theta\;\left[\frac{1}{2}\Phi\left(D^{2}+m\right)\Phi+\frac{\lambda}{6}\,\Phi^{3}-\frac{\lambda}{24}U\,(D^{2}\Phi)^{3}\right]\thinspace.\label{action superfield}
\end{align}
After we have written the action in this form, we can use the standard
tools of superspace perturbation theory. We start by using the background
method, splitting the superfield into its classical and quantum parts,
\begin{align}
\Phi\rightarrow\Phi+\Phi_{q} & \,,\label{background}
\end{align}
and then integrating over the quantum superfields $\Phi_{q}$ in the
path integral. The perturbative expansion involves the usual Feynman
rules of the three-dimensional WZ model, the propagator given by 
\begin{equation}
\left\langle \Phi\Phi\right\rangle =\frac{D^{2}-m}{k^{2}+m^{2}}\,\delta\left(\theta-\theta'\right)\,,\label{Propagator}
\end{equation}
the trilinear vertex factor corresponding to the coupling constant
$\lambda$, together with two additional vertices involving the spurion
$U$, represented in Figure\,\ref{fig:New-vertices}.

\begin{figure}[H]
\begin{centering}
\includegraphics[scale=0.45]{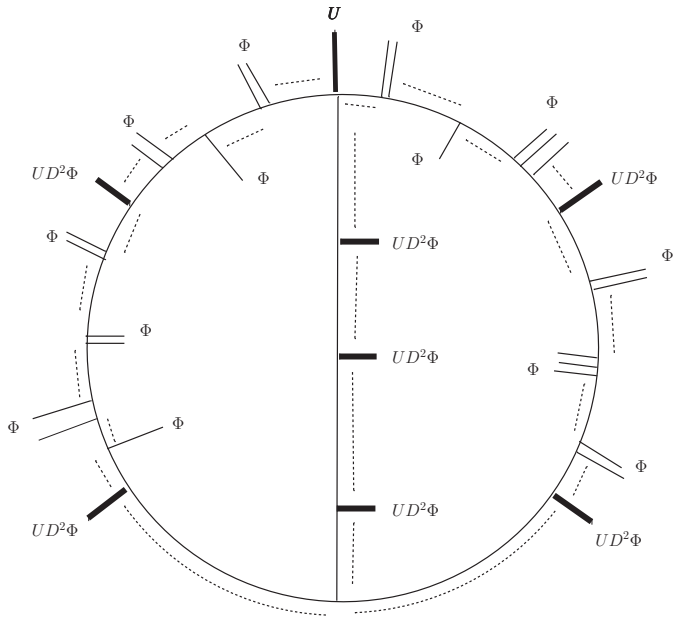}
\par\end{centering}
\caption{Two loop diagram with one insertion of $U$-vertex}

\label{Figure_1}
\end{figure}

\begin{figure}[H]
\begin{centering}
\includegraphics[scale=0.45]{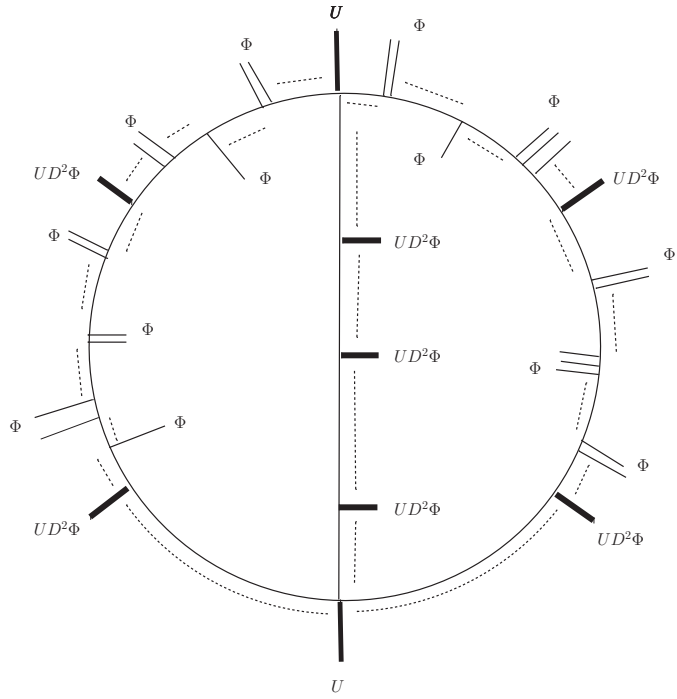}
\par\end{centering}
\caption{Two loop diagram with two insertion of $U$-vertex}

\label{Figure_2}
\end{figure}

\begin{figure}[h]
\begin{centering}
\includegraphics[scale=0.45]{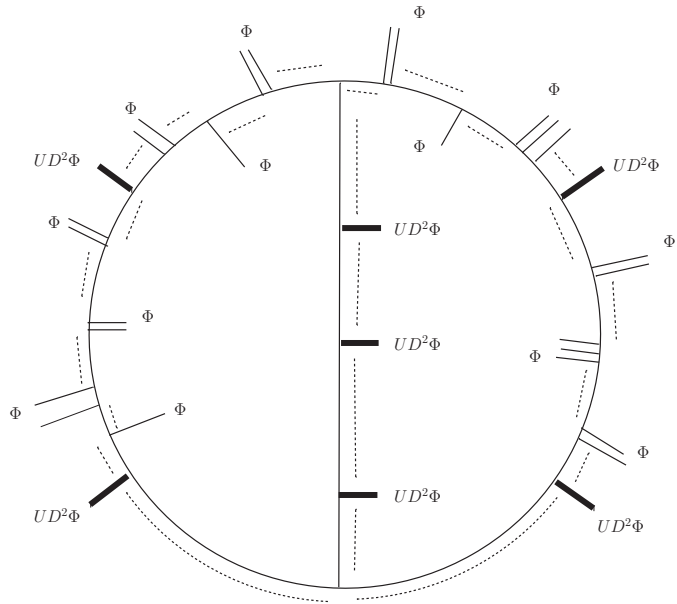}
\par\end{centering}
\caption{Two loop diagram without insertion of $U$-vertex}

\label{Figure_3}
\end{figure}

Using regularization by dimensional reduction\,\cite{Siegel:1979wq},
one loop diagrams are finite, so we study the possibly divergent diagrams
at the two-loop level. We follow the general strategy of\,\cite{Grisaru2003},
looking for the superficial degree of divergence of a general diagram
containing several possible insertion of $U$-vertices. We have to
consider three distinct classes of diagrams, represented in Figures\,\ref{Figure_1},
\ref{Figure_2} and \ref{Figure_3}, according to the number of quartic
vertices including the spurion.

We start with the class of two-loop graphs represented in Figure~\ref{Figure_1}.These
diagrams contains $p$ $U$-vertices and $k$ $\Phi^{3}$-vertices.
To study the possible divergent configurations we compute the mass
dimensions of the corresponding integrals once the $D$-algebra has
been performed, then we need to know the number of $D$'s and propagators
in the graph, always considering the most divergent configurations.
We take into account that we generate momentum factors through the
algebraic relations
\begin{equation}
\left(D^{2}\right)^{2}\text{=}\square\,,\thinspace\thinspace D^{2}D_{m}=P_{mb}\,D^{b}\,.\label{eq:D2momenta}
\end{equation}

The number of propagators and initial number of $D^{2}$ is calculated
as follows:
\begin{itemize}
\item Number of $D^{2}$ factors : $3p+k+3$, corresponding to

\begin{itemize}
\item 3 for each $U$-Vertex,
\item 2 for each $U\,D^{2}\Phi$-vertex,
\item 1 for each propagator $\left\langle \Phi\Phi\right\rangle $.
\end{itemize}
\item Number of propagators $\left\langle \Phi\Phi\right\rangle $: $k+p+2$.
\end{itemize}
Since the spurion $U$ has only the $\theta^{2}$ component, we have
to move a factor of $D^{2}$ onto $(p-1)$ $U$ factors to obtain
a final expression different from zero. We also use a factor of $D^{2}$
to contract each loop to a point. Then, the number of remaining factors
of $D^{2}$ will be $k+2p+2$. These $D^{2}$ factors will lead to
powers of momenta in the numerator according to Eq.\,\eqref{eq:D2momenta}.
Finally, taking into account the denominators of the propagators,
we end up with a final integrand of the general form
\begin{equation}
\int\,d^{6}q\:\frac{1}{q^{\,k+2}}\thinspace,\label{I_1}
\end{equation}
which by power counting is divergent if $k\leq4$. This condition
does not depend of $p$, therefore for each fixed $k$ we have an
arbitrary number $p$ of $U$ insertions, all of them being in principle
superficially divergent. That means the model has an infinite number
of potentially divergent diagrams, which indicates non-renormalizability.
This result is different from that in four dimensions\,\cite{Grisaru2003},
where the value of $p$ which could yield a divergent diagram was
bounded, so the number of potentially divergent diagrams was finite.
The notion of chirality, essential to four-dimensional SUSY, imposes
additional conditions on the manipulation of covariant superderivatives,
such that actually one can just have a divergent diagram with $p\leq1$,
i.e., with at the most a single $U$ insertion.

\medskip{}

For the diagrams represented in Figure\,\ref{Figure_2}, the number
of propagators and initial $D^{2}$ factors are, respectively, $k+p+1$
and $3p+k+3$. Then one can conclude these diagrams will behave like
\begin{equation}
\int\,d^{6}q\:\frac{1}{q^{\,k}},
\end{equation}
which will be divergent if $k\leq6$, again irregardless of $p$.
Finally, for diagrams like the one depicted in Figure\,\ref{Figure_3},
we start with $3p+k+3$ $D^{2}$ factors and $k+p+3$ propagators,
and the final integrand ends up being of the form
\begin{equation}
\int\,d^{6}q\:\frac{1}{q^{\,k+4}},
\end{equation}
which will be divergent if $k\leq2$, again for arbitrary number of
$U$ insertions.

These results indicate that, unless some unexpected cancellation or
additional constraints can be shown to occur in the infinite number
of potentially divergent diagrams, we have a non-renormalizable model.
This is in contrast with the results found in ${\cal N}=1/2$ models\,\cite{Grisaru:2003fd,Grisaru:2004qw}
in four dimensions, despite the fact that in our case the final effect
of the deformation in the classical action is the inclusion of a single
additional interaction term, as in those papers. Despite the similarities
in the diagrammatic expansion of both cases, the existence of the
notion of chirality in four space-time dimensions imposes additional
constraints in the power counting of the model, which contributes
to ensure renormalizability. Finally, it is also interesting to remark
that WZ models in $\left(3+1\right)$ dimensions deformed by means
of a Drinfel'd twist were also shown to be non-renormalizable\,\cite{Dimitrijevic:2007cu,Dimitrijevic2009,Dimitrijevic2011,Dimitrijevic:2010yv},
similarly to what we found in our model.

\section{\label{sec:Concluding-Remarks}Concluding Remarks}

The deformation of supersymmetric models using the concept of a Drinfel'd
twist preserves several important algebraic properties of the SUSY
algebra, and it could allow for the definition of deformed supersymmetric
models with interesting properties. However, when trying to put the
Hopf algebraic formalism to work in a specific physical model, one
often encounters difficulties. In\,\cite{Dimitrijevic:2007cu,Dimitrijevic:2011zg},
for example, a particular twist in four space-time dimensions was
considered, and it was shown that even a formally supersymmetric covariant
action involving superfields could fail to be SUSY invariant, when
projected in terms of the component fields. In this case, the notion
of chirality seems to be responsible for these problems, since it
forces one to introduce non-local projection operators in the formalism.
Also, a simple generalization of the WZ model failed to be renormalizable. 

In three space-time dimensions, there is no notion of chirality, so
in principle the application of the Drinfel'd twist would be simpler,
and could open up the possibility of studying deformed ${\cal N}=1$
supersymmetric models, which is difficult to do in other formalisms.
Therefore, we studied a twist deformation of three-dimensional ${\cal N}=1$
SUSY, and defined what would be the simplest non-anticommutative WZ
model in this context. However, we showed that this theory suffers
from the same problems present in the four-dimensional case. At the
classical level, the model fails to be invariant under deformed SUSY
transformations, meaning that although the star product and the deformed
coproduct are algebraically compatible, this compatibility does not
guarantee actual SUSY invariance of the physical model under consideration.
At the quantum level, the WZ model is finite at the one loop level,
but at two loops there are an infinite number of potentially divergent
diagrams, rendering the model non-renormalizable.

Our results reinforces the idea that algebraic consistency doest not,
by itself, guarantee the definition of physically meaningful deformed
models, either at the classical or at quantum level. We studied a
very simple twist deformation, and one might conjecture whether there
are more complicated twists that could lead to consistent theories.
This is a problem that we leave for future studies.

\bigskip{}

\textbf{Acknowledgements.} This work was supported by Conselho Nacional
de Desenvolvimento Científico e Tecnológico (CNPq), Fundação de Amparo
a Pesquisa do Estado de São Paulo (FAPESP) and Coordenação de Aperfeiçoamento
de Pessoal de Nível Superior (CAPES), via the following grants: CNPq
482874/2013-9, FAPESP 2013/22079-8 and 2014/24672-0 (AFF), CAPES PhD
grant (AGQ and CP).

\end{document}